\documentclass[prb,aps,twocolumn,superscriptaddress]{revtex4}

\usepackage{amsmath}
\usepackage{amsfonts}
\usepackage{bm}
\usepackage{graphicx}

\begin{document}

\title{Self-rotation and synchronization in exciton-polariton
condensates}

\author{Hiroki Saito}
\affiliation{Department of Engineering Science, University of
Electro-Communications, Tokyo 182-8585, Japan}

\author{Rina Kanamoto}
\affiliation{Department of Physics, Meiji University, Kawasaki, Kanagawa
214-8571, Japan}

\date{\today}

\begin{abstract}
Self-rotation occurs in an exciton-polariton condensate in a
two-dimensional semiconductor microcavity pumped by a nonresonant Gaussian
laser beam.
A wave packet of the condensate spontaneously rotates around the center
of the pumped region at a constant frequency breaking the rotation
symmetry of the system.
When two self-rotating condensates are created with an appropriate
distance, synchronization occurs between the dynamics of the self-rotating
condensates.
\end{abstract}


\maketitle

\section{Introduction}

Self-oscillations are ubiquitous phenomena that occur in nature, ranging
from whistles and heartbeats to pulsating variable stars~\cite{Jenkins}.
Self-oscillation is induced by some instability and maintained by the
oscillation itself, i.e., external energy is synchronously taken into the
system, which balances with dissipation, and results in the stable limit
cycle in the phase space.
When two or more self-oscillators are coupled with each other,
synchronization can occur between their oscillations, as Huygens observed
in pendulum clocks~\cite{Huygens}.

Self-oscillation can be observed in open quantum systems with
nonlinearity, in which energy is replenished from and dissipated into
their environments.
Laser and Josephson oscillations may be regarded as self-oscillations in
quantum systems.
Self-oscillations have also been observed in optomechanical
systems~\cite{Carmon,Kippenberg} and in a resonator driven by a
superconducting single-electron transistor~\cite{Rodriques}.
Here, we focus on exciton-polariton condensates in a planar
microcavity~\cite{Carusotto,Byrnes} as an open quantum system to
investigate the self-oscillation.
Exciton-polaritons excited by external laser beams have a lifetime of
picoseconds, which is of the same order as the characteristic time scale
of polariton dynamics.
Therefore, the exciton-polariton condensate is a nonequilibrium open
system.
Various experiments have been performed on this system, such as
observations of Bose-Einstein condensation~\cite{Kasprzak}, quantized
vortices~\cite{Lagoudakis,Roumpos}, quantum
hydrodynamics~\cite{Amo09,Nardin,Amo11,Grosso},
and a chaotic non-Hermitian billiard~\cite{Gao}.
Self-oscillations and synchronization in exciton-polariton condensates
have been predicted in
Refs.~\onlinecite{Wouters08,Eastham,Borgh,Read,Bob,Rayanov,Ma,Chen},
and the phase locking of condensates has been reported in
Ref.~\onlinecite{Baas,Lagoudakis2}.
The B\'enard-von K\'arm\'an vortex street in an exciton-polariton
superfluid~\cite{Saito} can also be regarded as self-oscillation.

In this paper, we show that a wave packet of an exciton-polariton
condensate that is nonresonantly pumped by a Gaussian laser beam rotates
spontaneously around the center of the pumped region.
This self-rotation is a consequence of spontaneous breaking of 
the rotation and clockwise-counterclockwise symmetries, and the system
acquires an angular momentum.
We also show that when two self-rotating condensates are excited by two
Gaussian laser beams with an appropriate distance between them, the
rotating dynamics are synchronized, even when the individual rotation
frequency is different.

This paper is organized as follows.
The problem is formulated and the properties of the uniform system are
reviewed in Sec.~\ref{s:uniform}.
The self-rotation of the condensate is investigated in
Sec.~\ref{s:rotation}.
The synchronization of two self-rotating condensates is demonstrated in
Sec.~\ref{s:synchro}.
A conclusion and discussion are provided in Sec.~\ref{s:conc}.

\section{Modulational instability in a homogeneous system}
\label{s:uniform}

First, we formulate the problem and give a brief review of the
modulational instability in a homogeneous system.
The dynamics of an exciton-polariton condensate with an effective mass
$m$ is modeled by the nonlinear Schr\"odinger equation given
by~\cite{Wouters07}
\begin{equation} \label{GP1}
i\hbar \frac{\partial \psi}{\partial t} = -\frac{\hbar^2}{2m} \nabla^2
\psi + \frac{i\hbar}{2} (R n_R - \gamma_c) \psi + g_c |\psi|^2 \psi + g_R
n_R \psi,
\end{equation}
where $\psi(\bm{r}, t)$ is the macroscopic wave function of the
exciton-polariton condensate and $n_R(\bm{r}, t)$ represents the polariton
density of the reservoir, which obeys
\begin{equation} \label{GP2}
\frac{\partial n_R}{\partial t} = P - \gamma_R n_R - R n_R |\psi|^2.
\end{equation}
The external laser pump $P$ creates polaritons with much larger
energy than that of the condensate polaritons in the lower branch.
The pumped polaritons then relax and form the reservoir $n_R$, which feeds
polaritons to the condensate with a rate $R$.
In this section, we consider a homogeneous pump with a constant $P$.
The loss rates of the condensate and reservoir polaritons are $\gamma_c$
and $\gamma_R$, respectively.
In Eq.~(\ref{GP1}), the condensate polaritons interact with each other
through the nonlinear term with a coefficient $g_c$, and the interaction
between the condensate and reservoir polaritons is described by the term
with $g_R$.

\begin{figure}[tbp]
\includegraphics[width=8cm]{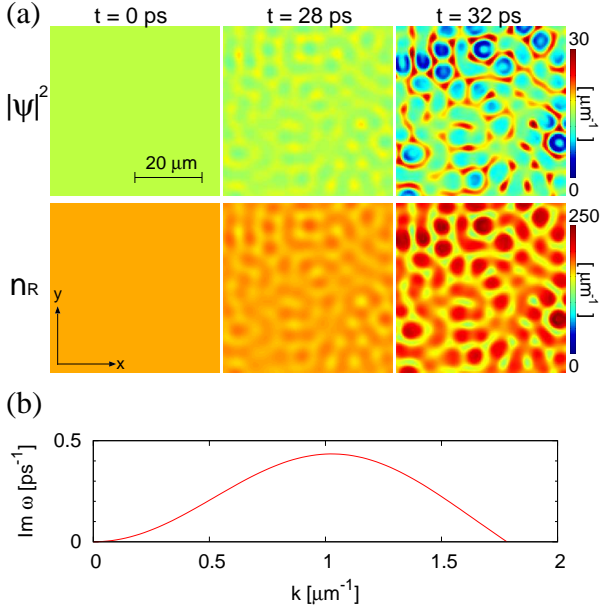}
\caption{
(color online) Modulational instability in the uniform system with
$\gamma_c^{-1} = 0.15$ ps, $\gamma_R^{-1} = 3$ ps, $g_c = 6 \times
10^{-3}$ ${\rm meV} \cdot \mu{\rm m}^2$, $g_R = 2 g_c$, $R = 0.04$
${\rm ps}^{-1} \mu{\rm m}^2$, $P = 160$ ${\rm ps}^{-1} \mu{\rm m}^{-2}
\simeq 2.9 P_{\rm th}$, and $m = 5 \times 10^{-5}$ electron mass.
(a) Time evolution of the condensate and reservoir density profiles,
$|\psi|^2$ and $n_R$, where the initial state is the stationary state.
(b) Positive imaginary part of the Bogoliubov spectrum.
}
\label{f:uniform}
\end{figure}
In a homogeneous system, the polariton densities in the steady state above
the condensation threshold are given by $|\psi_0|^2 = (P - P_{\rm th}) /
\gamma_c$ and $n_{R0} = \gamma_c / R$, where $P_{\rm th} = \gamma_c
\gamma_R / R$.
This homogeneous solution is modulationally unstable for a weak pumping
regime~\cite{Wouters07,Bob,Smirnov,Liew}, which
can be explained by the effective attractive interaction in the
condensate~\cite{Smirnov}.
Figure~\ref{f:uniform}(a) shows the dynamics of the density profiles under
modulational instability, where the initial state is $\psi = \psi_0$
and $n_R = n_{R0}$ with a small random noise, which is a random number set
on each numerical mesh.
In this and the following numerical simulations, the pseudospectral method
with the fourth-order Runge-Kutta scheme is employed.
The unstable wavelengths can be obtained from the Bogoliubov analysis.
Substituting $\psi(\bm{r}, t) = [\psi_0 + u_k
e^{i \bm{k} \cdot \bm{r} - i \omega t} + v_k^*
e^{-i \bm{k} \cdot \bm{r} - i \omega^* t}] e^{-i \mu t / \hbar}$
and $n_R(\bm{r}, t) = n_{R0} + {\rm Re}(n_k e^{i \bm{k} \cdot \bm{r}
- i \omega t})$ with $\mu = g_c |\psi_0|^2 + g_R n_{R0}$ into
Eqs.~(\ref{GP1}) and (\ref{GP2}) and diagonalizing the $3 \times 3$
matrix, we obtain the Bogoliubov spectrum $\omega(k)$ and mode
coefficients $u_k$, $v_k$, and $n_k$.
When the imaginary part of $\omega(k)$ is positive, the mode with the
wave number $k$ is unstable.
Figure~\ref{f:uniform}(b) shows the imaginary part of $\omega(k)$.
The most unstable wavelength is $\lambda \simeq 2\pi$ $\mu {\rm m}$, which
agrees with the pattern in Fig.~\ref{f:uniform}(a).

\section{Spontaneous rotation with a Gaussian pump}
\label{s:rotation}

We consider a situation in which exciton-polaritons are pumped by a
Gaussian laser beam.
The pump profile is given by
\begin{equation}
P(\bm{r}) = P_0 e^{-r^2 / \rho^2},
\end{equation}
where $P_0$ is the peak intensity and $\rho$ is the $1/e$ width of the
beam.
When the pump width $\rho$ is much larger than the unstable wavelengths of
a homogeneous system pumped by $P = P_0$, the central region of the pump
is sufficiently wide for modulational instability and pattern formation.
On the other hand, when $\rho$ is much smaller than the unstable
wavelengths, no modulational instability occurs in the Gaussian-shaped
condensate.
An interesting situation is expected for $\rho$ comparable to the unstable
wavelength.
Self-oscillations have been observed in a one-dimensional system in this
regime~\cite{Bob}.

\begin{figure}[tbp]
\includegraphics[width=8cm]{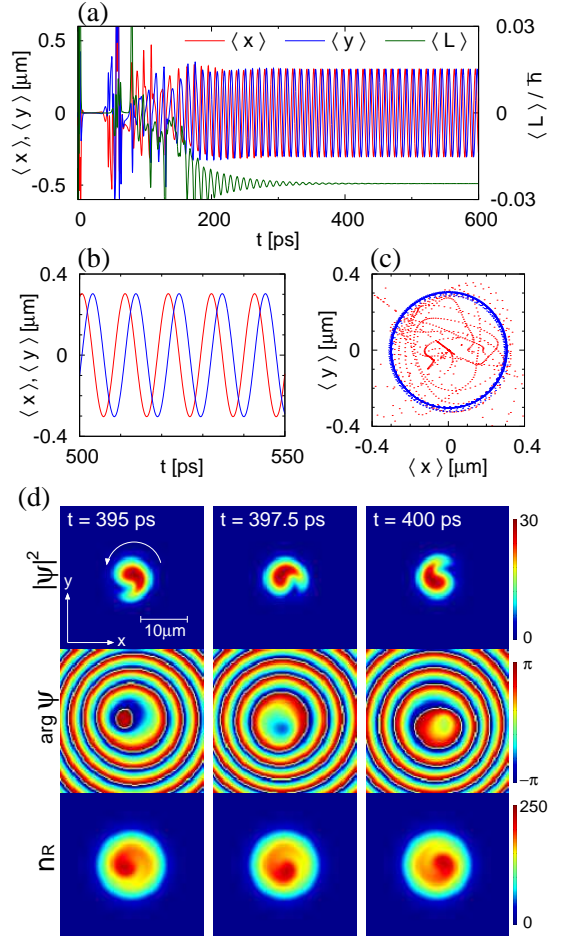}
\caption{
(color online) Time evolution of the system pumped by a Gaussian laser
beam with $P_0 = 160$ ${\rm ps}^{-1} \mu{\rm m}^{-2}$
and $\rho = 5$ $\mu {\rm m}$.
The other parameters are the same as those used in Fig.~\ref{f:uniform}.
(a) Time evolution of the center-of-mass (COM) position $(\langle x
\rangle, \langle y \rangle)$ and the angular momentum $\langle L \rangle$
of the condensate.
(b) Magnification of (a).
(c) Trajectory of $(\langle x \rangle, \langle y \rangle)$, where the red
and blue points are $t < 200$ ps and $t > 200$ ps, respectively.
The points are plotted every 0.17 ps.
(d) Snapshots of the density and phase profiles, $|\psi|^2$, ${\rm arg}
\psi$, and $n_R$.
The white arrow indicates the direction of rotation for the COM position.
See the Supplemental Material for a movie of the condensate
dynamics~\cite{SM}.
}
\label{f:rotation}
\end{figure}
Figure~\ref{f:rotation} shows the dynamics of the system pumped by a
Gaussian laser beam with $\rho = 5$ $\mu{\rm m}$, where the initial state
is $\psi = n_R = 0$ plus a small random noise.
Figures~\ref{f:rotation}(a)-\ref{f:rotation}(c) show the time evolution of
the center-of-mass (COM) position,
\begin{equation}
\langle \bm{r} \rangle = \frac{\int \bm{r} |\psi|^2 d\bm{r}}{\int |\psi|^2
d\bm{r}},
\end{equation}
and the angular momentum per particle,
\begin{equation}
\langle L \rangle = \frac{-i \hbar \int \psi^* (x \partial_y - y
\partial_x) \psi d\bm{r}}{\int |\psi|^2 d\bm{r}},
\end{equation}
of the condensate.
After $t \sim 200$ ps, the COM motion becomes periodic and the angular
momentum converges to a constant value.
The trajectory of $\langle \bm{r} \rangle$ is entrained into a circle and
rotates along a circle at a constant angular frequency.
This circular trajectory is robust against external perturbation.
We have confirmed that the trajectory always returns to the circle after
the regular dynamics are disturbed by a temporal external potential.

Figure~\ref{f:rotation}(d) shows the time evolution of the density and
phase profiles for the regular motion.
A tadpole-shaped condensate wave packet rotates around the center
counterclockwise.
It is interesting to note that the angular momentum $\langle L \rangle$ is
negative, whereas the tadpole-shaped wave packet rotates counterclockwise.
This seeming contradiction arises from the superflow in the condensate.
Figure~\ref{f:rotation}(d) shows ${\rm arg} \psi$, which indicates that
the superfluid flows out from the point near the head of the tadpole,
and results in superflow from the head to the tail of the tadpole.
This is because the condensate is fed from the reservoir mainly at the
head of the tadpole, because the density of the reservoir is large ahead
of the tadpole.
As a result, the tadpole grows ahead, while the tail is always shrunk by
the dissipation.
This is the reason why the directions of the tadpole rotation and the
angular momentum are opposite.
For initial noises with different random seeds, $\langle L \rangle > 0$
and $\langle L \rangle < 0$ occurs with the same probability due to the
chiral symmetry of the system.

The pump with a Gaussian beam thus induces the rotation dynamics, which
can be regarded as self-rotation, because the rotation symmetry of the
system is spontaneously broken and the stable rotation persists without
any external rotating force.
The rotation is driven by the interplay between the condensate and
reservoir; the tadpole-shaped condensate grows ahead due to the
density gradient in the reservoir, while the high density region of the
reservoir chases the tail of the tadpole where the supply to the
condensate is small.

\begin{figure}[t]
\includegraphics[width=8cm]{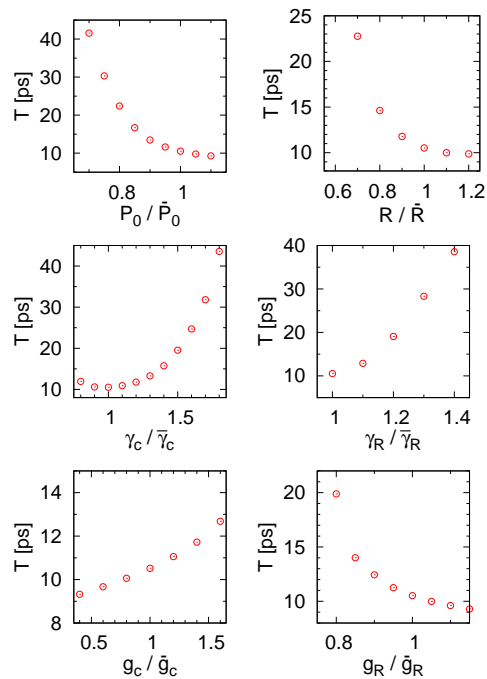}
\caption{
(color online) Parameter dependence of the period $T$ of the
self-rotation.
The parameters used in Fig.~\ref{f:rotation} are denoted as $\bar{P}_0$,
$\bar{R}$, $\bar{\gamma}_c$, $\bar{\gamma}_R$, $\bar{g}_c$, and
$\bar{g}_R$.
}
\label{f:depend}
\end{figure}
Figure~\ref{f:depend} shows the parameter dependence of the period of the
self-rotation.
The self-rotation frequency increases with $P_0$, $R$, and $g_R$,
and decreases as $\gamma_c$, $\gamma_R$, and $g_c$ increase.

\section{Synchronization between two rotating condensates}
\label{s:synchro}

We study the case in which two self-rotating condensates are excited by
two Gaussian laser beams, and examine whether synchronization occurs
between the two self-rotations.
The external pump is given by
\begin{equation} \label{twogauss}
P(\bm{r}) = P_{\rm left} e^{-(\bm{r} - \bm{R}_{\rm left})^2 / \rho^2}
+ P_{\rm right} e^{-(\bm{r} - \bm{R}_{\rm right})^2 / \rho^2},
\end{equation}
where $P_{\rm left, right}$ and $\bm{R}_{\rm left, right} =
(X_{\rm left, right}, Y_{\rm left, right})$ are the peak intensities and
the positions of the two Gaussian-shaped pumps, respectively.
Without loss of generality, we set $Y_{\rm left, right} = 0$.
The pump intensities are taken to be $P_{\rm left} = 160$ ${\rm ps}^{-1}
\mu{\rm m}^{-2}$ and $P_{\rm right} = 0.99 P_{\rm left}$, which results in
different self-rotation frequencies (see Fig.~\ref{f:depend}).
If the two self-rotating condensates are independent (i.e. in the case of
$X_{\rm right} - X_{\rm left} \rightarrow \infty$), then the rotation
period is 10.52 ps for $P_{\rm left}$ and 10.70 ps for $P_{\rm right}$.
The COM position of each condensate is defined relative to the center of
the pump as
\begin{equation}
\langle \bm{r} \rangle_{\rm left, right} =
\frac{\int_{|\bm{r} - \bm{R}_{\rm left, right}| < r_c}
(\bm{r} - \bm{R}_{\rm left, right}) |\psi|^2 d\bm{r}}
{\int_{|\bm{r} - \bm{R}_{\rm left, right}| < r_c} |\psi|^2 d\bm{r}},
\end{equation}
where the cutoff radius $r_c$ is taken to be 8 $\mu{\rm m}$, which is
sufficient to cover each condensate.

\begin{figure}[tbp]
\includegraphics[width=8cm]{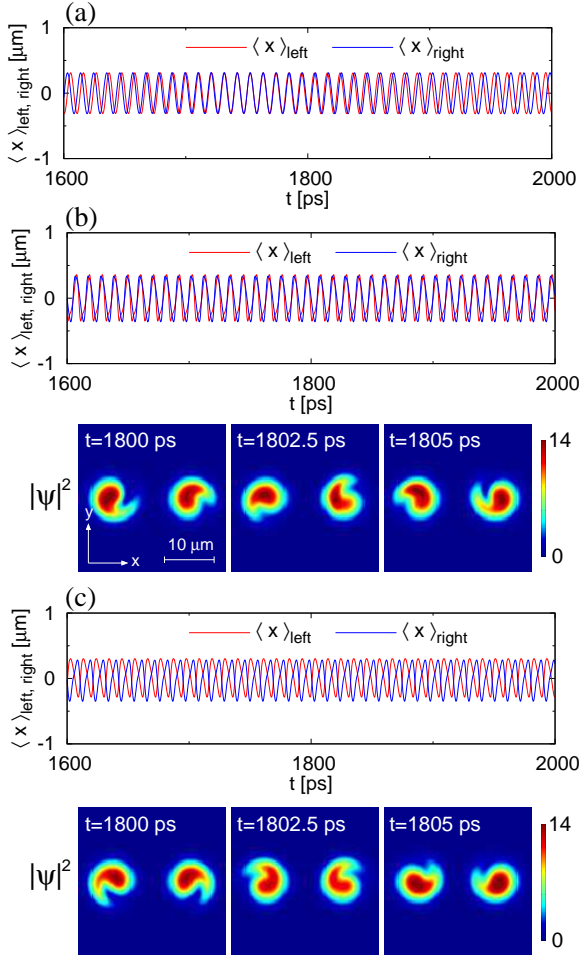}
\caption{
(color online) Dynamics of two self-rotating condensates pumped according
to Eq.~(\ref{twogauss}) with $P_{\rm left} = 160$
${\rm ps}^{-1} \mu{\rm m}^{-2}$ and $P_{\rm right} = 0.99 P_{\rm left}$.
The other parameters are the same as those in Fig.~\ref{f:rotation}.
(a) Time evolution of $\langle x \rangle_{\rm left}$ and
$\langle x \rangle_{\rm right}$ with $X_{\rm right} - X_{\rm left} =
20$ $\mu{\rm m}$.
The two self-rotations are not synchronized.
(b) Time evolution of $\langle x \rangle_{\rm left}$ and
$\langle x \rangle_{\rm right}$ (upper panel) and snapshots of the density
profile $|\psi|^2$ (lower panels), where $X_{\rm right} - X_{\rm left} =
16$ $\mu{\rm m}$.
Synchronization occurs between the two self-rotating condensates.
(c) The same conditions as those in (b), except a random number seed used
to produce the initial noise.
See the Supplemental Material for movies of the dynamics in (b) and
(c)~\cite{SM}.
}
\label{f:synchro}
\end{figure}
Figure~\ref{f:synchro}(a) shows the time evolution of the two
self-rotating condensates, where the distance between the two pumps is
$X_{\rm right} - X_{\rm left} = 20$ $\mu{\rm m}$.
In this case, the two condensates are well separated from each other and
the interaction between them is negligible.
In Fig.~\ref{f:synchro}(a), the two self-rotations are in opposite
directions, while they also rotate in the same direction depending on the
initial noise.
The left- and right-side condensates rotate independently with periods of
10.52 and 10.70 ps, respectively, and no synchronization occurs.
Figure~\ref{f:synchro}(b) shows the synchronization for $X_{\rm right} -
X_{\rm left} = 16$ $\mu{\rm m}$.
The COM positions $\langle x \rangle_{\rm left}$ and $\langle x
\rangle_{\rm right}$ oscillate in phase at the same frequency.
The two condensates rotate in opposite directions with a common rotation
period of 10.53 ps, by which it seems that the right-side frequency is
entrained into the left-side frequency~\cite{Pikovsky}.
Interestingly, another type of synchronization is observed for the same
parameters, which is shown in Fig.~\ref{f:synchro}(c).
The COM positions $\langle x \rangle_{\rm left}$ and $\langle x
\rangle_{\rm right}$ oscillate out of phase with a period of 10.55 ps.
The only difference between the numerical simulations in
Fig.~\ref{f:synchro}(b) and \ref{f:synchro}(c) is the random number
seed used to produce the small initial noise.

\begin{figure}[tbp]
\includegraphics[width=8cm]{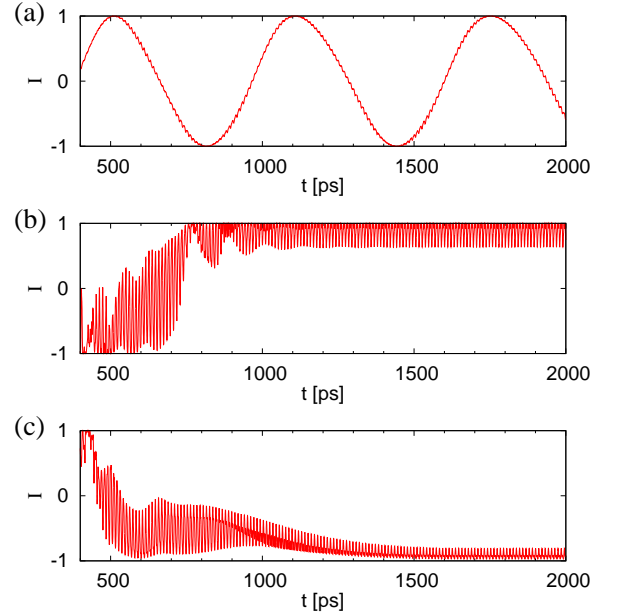}
\caption{
(color online) Time evolution of $I$ defined in Eq.~(\ref{inner}), where
(a)-(c) correspond to the dynamics shown in
Figs.~\ref{f:synchro}(a)-\ref{f:synchro}(c).
}
\label{f:inner}
\end{figure}
To clearly visualize the synchronization, the inner product is calculated:
\begin{equation} \label{inner}
I = \hat x_{\rm left} \hat x_{\rm right} - \hat y_{\rm left}
\hat y_{\rm right},
\end{equation}
where $\hat{\bm{r}}_{\rm left, right} = \langle \bm{r}
\rangle_{\rm left, right} / |\langle \bm{r} \rangle_{\rm left, right}|$
are the unit vectors of the COM positions relative to the centers of the
pumps.
The minus sign in Eq.~(\ref{inner}) is introduced to quantify the
synchronization between the opposite rotations, as shown in
Figs.~\ref{f:synchro}(b) and \ref{f:synchro}(c).
If the two unit vectors rotate in the opposite directions as $x_{\rm left}
+ i y_{\rm left} = e^{i \omega t}$ and $x_{\rm right} + i y_{\rm right} =
e^{-i \omega' t + \phi}$, then $I = \cos[(\omega - \omega')t + \phi]$.
When frequency lock occurs, $I$ becomes constant.

Figures~\ref{f:inner}(a)-\ref{f:inner}(c) show the time evolution of $I$,
which correspond to the dynamics shown in
Figs.~\ref{f:synchro}(a)-\ref{f:synchro}(c).
When the two self-rotations are not synchronized, $I$ oscillates at their
frequency difference, as shown in Fig.~\ref{f:inner}(a), where the period
is $\simeq (10.52^{-1} - 10.7^{-1})^{-1} \simeq 625$ ps.
When synchronization occurs, $I$ oscillates around a constant value,
as shown in Figs.~\ref{f:inner}(b) and \ref{f:inner}(c).
The synchronized phase $\phi$ is $\simeq 0$ in Fig.~\ref{f:inner}(b) and
$\simeq \pi$ in Fig.~\ref{f:inner}(c).

\section{Conclusions and Discussion}
\label{s:conc}

We have investigated the dynamics of an exciton-polariton condensate in a
planar semiconductor microcavity pumped by a nonresonant laser beam.
We considered a parameter region in which modulational instability arises
in a homogeneous system (Fig.~\ref{f:uniform}).
It was shown that when the polariton condensate is excited by a Gaussian
laser beam of which the width is comparable to the unstable wavelength,
the self-rotation of the condensate occurs; the COM position of the
condensate wave packet rotates around the center of the pump at a
constant frequency (Fig.~\ref{f:rotation}).
The parameter dependence of the self-rotation frequency was investigated
(Fig.~\ref{f:depend}).
In addition, the synchronization of two self-rotating condensates pumped
by two Gaussian laser beams was investigated (Figs.~\ref{f:synchro} and
\ref{f:inner}).
The two self-rotations could be synchronized, even when the individual
rotation frequencies were slightly different.

Experimentally, there always exist imperfections in a sample of the
semiconductor microcavity, which produces random potential for polaritons.
We have performed numerical simulations including random potential with
a typical energy of $\sim 0.1$ meV and spatial scale of $\sim 10$
$\mu{\rm m}$, and confirmed that the self-rotation and the synchronization
occur even in the presence of the disorder potential.

Although it was assumed that $\gamma_c \gg \gamma_R$ in this work, the
loss rates satisfy $\gamma_c \lesssim \gamma_R$ for typical samples in the
experiments.
The condition $\gamma_c \gg \gamma_R$ may be satisfied by a reduction of
the Q-factor of the microcavity at the frequency of the condensate
polariton, which may be realized near the edge of the stopband of the
distributed Bragg reflector.

Spontaneous rotation has been studied for the nonlinear Schr\"odinger
equation with pump and decay~\cite{Keeling}, and recently a wave packet of
a condensate in a harmonic potential was found to rotate
spontaneously~\cite{Berman}.
The mechanism for these spontaneous rotations may be different from that
presented here, because only the condensate wave function was considered
in these previous studies, while the interplay between the condensate and
the reservoir plays an important role in the self-rotation in this work.

\begin{acknowledgments}
The authors thank S. Kaneko for his participation in the early stage of
this work.
This work was supported by JSPS KAKENHI Grant Numbers JP16K05505, 
JP26400414, and JP25103007.
\end{acknowledgments}

\end{document}